\documentclass[twocolumn,amsmath,amssymb,floats,superscriptaddress,nofootinbib,10pt]{revtex4-1}

\pdfoutput=1

\AtBeginDocument{%
    \newwrite\bibnotes
    \def\bibnotesext{Notes.bib}
    \immediate\openout\bibnotes=\jobname\bibnotesext
    \immediate\write\bibnotes{@CONTROL{REVTEX41Control}}
    \immediate\write\bibnotes{@CONTROL{%
    apsrev41Control,author="08",editor="1",pages="1",title="0",year="1"}}
     \if@filesw
     \immediate\write\@auxout{\string\citation{apsrev41Control}}%
    \fi
}%

\usepackage{graphicx}
\usepackage{dcolumn}
\usepackage{bm}
\usepackage{revsymb}
\usepackage[usenames]{color}
\usepackage{color}
\usepackage{physics}
\usepackage{amsmath}
\usepackage[usenames]{color}
\usepackage{amsfonts}
\usepackage{graphicx}
\usepackage{dcolumn}
\usepackage{bm}
\usepackage{revsymb}
\usepackage[usenames]{color}
\usepackage{graphicx}

\usepackage{color}
\usepackage{physics}
\usepackage{amsmath}
\usepackage{amssymb,bbm}
\usepackage{subfig}

\usepackage{xcolor}

\newcommand{\bea}{\begin{eqnarray}}
\newcommand{\eea}{\end{eqnarray}}

\definecolor{nblue}{RGB}{28,130,185}
\definecolor{cgreen}{RGB}{76,153,0}
\definecolor{myorange}{RGB}{245,156,74}

\usepackage{hyperref}
\hypersetup{
  colorlinks=true,
  citecolor=magenta,
  urlcolor=-myorange
}

\usepackage{xcolor}
\definecolor{ogreen} {RGB}{71,191,145}
\newcommand{\rl}[1]{\textcolor{ogreen}{#1}}

\usepackage{hyperref}
\hypersetup{
  colorlinks=true,
  citecolor=magenta,
  urlcolor=-myorange
}

\begin{document}

\title{Odd viscous flow past a sphere at low but nonzero Reynolds numbers}

\author{Ruben Lier}
\email{r.lier@uva.nl}
\affiliation{Institute for Theoretical Physics, University of Amsterdam, 1090 GL Amsterdam, The Netherlands}
\affiliation{Dutch Institute for Emergent Phenomena (DIEP), University of Amsterdam, 1090 GL Amsterdam, The Netherlands}
\affiliation{Institute for Advanced Study, University of Amsterdam, Oude Turfmarkt 147, 1012 GC Amsterdam, The Netherlands}

\begin{abstract}
Measuring lift force on symmetrically shaped obstacles immersed in laminar flow is the quintessential way of signalling odd viscosity. For flow past cylinders, such a lift force does not arise when incompressibility and no-slip boundary conditions hold, whereas for spheres, a lift force was found in Stokes flow, applying to cases where the Reynolds number is negligible and convection can be ignored. When considering the role of convection at low but nonzero Reynolds numbers, two arising hurdles are the Whitehead paradox and the breaking of axial symmetry, which are overcome by asymptotic matching and the Lorentz reciprocal theorem respectively. We also consider the case where axial symmetry is retained because the translation of the sphere is aligned with the anisotropy vector of odd viscosity. We find that while lift vanishes, the interplay between odd viscosity and convection gives rise to a stream-induced non-vanishing torque. 
\end{abstract}

\maketitle

\tableofcontents

\section{Introduction}
In the ever ongoing study of fluids, the exploration of 
flow instigated by non-dissipative, parity-breaking viscosities has emerged as a new frontier. Such transport coefficients, which fall under the heading of \emph{odd viscosity}, result from the spinning of microscopic or mesoscopic particles constituting the fluid \cite{Banerjee2017,markovich2021odd,Fruchart_2023}. Odd viscosity gives rise to a wide range of novel physical phenomena, such as azimuthal flow of sedimenting particles \cite{Khain_2022}, torque induced by the rate of area change \cite{PhysRevE.89.043019}, non-reciprocal self-induced flow fields for swimmers \cite{hosaka2023hydrodynamics}, formation of inertial-like waves \cite{Kirkinis_Olvera}, Hall-like transport in fluids \cite{Lou2022}, oscillating vortical boundary layers \cite{PhysRevFluids.5.104802} and many more \cite{PhysRevE.90.063005}. Furthermore, odd viscosity has implications across various scales and disciplines, from electron fluids \cite{berdyugin2019measuring,Delacr_taz_2017}, diatomic gases \cite{KORVING19665,HULSMAN197053}, equatorial waves \cite{tauber2019bulkinterface,souslov2019topological}, colloidal rotors \cite{soni2019odd,han2021fluctuating,hargus2020time,PhysRevLett.130.158201,PhysRevLett.127.178001,matus2024molecular} and biological systems \cite{Tan2022}.
\newline 
Odd viscosity can be found in multiple dimensions. Originally treated as a viscous analogue of two-dimensional Hall conductivity \cite{avron1995viscosity,avron1998odd,levay1995berry,hoyosreview,PhysRevLett.108.066805}, it was found that three dimensional fluids can display several odd viscosities \cite{Khain_2022}, provided that there is a preferred vector giving rise to anisotropy. Odd viscosity is also studied at different \emph{Reynolds numbers}, ranging from zero Reynolds number Stokes flow \cite{Khain_2022,olveira,hosaka2023lorentz} or small Reynolds number Oseen flow \cite{lier2022lift} to large Reynolds number flows with fully developed turbulence \cite{dewit2023pattern}. 
 \newline
A key tool for understanding fluid behavior at low Reynolds numbers is the \emph{Lorentz reciprocal theorem} \cite{lorentzoriginal}. However, this theorem assumes reciprocity, which the odd viscous tensor violates. This makes the Lorentz reciprocal theorem in its original form unusable for fluid systems with non-vanishing odd viscosity. Fortunately, it turns out that a generalized version of the Lorentz reciprocal theorem exists which takes this non-reciprocity into account by considering an auxiliary fluid system with an odd viscosity of opposite sign \cite{hosaka2023lorentz}.
\newline
One of the many things that this generalized Lorentz reciprocal theorem allows one to compute is the odd viscosity induced \emph{lift force} on a translating object. In two dimensions, such a lift force is zero for incompressible fluids with no-slip boundary conditions, but can get a non-vanishing value at quadratic order in slip length \cite{lier2024slipinduced}, for compressible fluids \cite{lier2022lift,hosaka2021nonreciprocal,PhysRevE.109.044126}, or liquid domains that lie within a membrane \cite{PhysRevE.104.064613}.
\newline 
In Refs.~\cite{PhysRevLett.132.218303,Khain_2022,hosaka2023lorentz}, the force and torque for odd Stokes flow past a sphere are computed. Stokes flow, also known as creeping flow, means that the role of inertia, which enters through the convective term in the Navier-Stokes equation, is ignored entirely. The reason why it is worthwhile to not ignore this convective terms is twofold. Firstly, to ignore the convective term requires the Reynolds number to be negligible, which is not the case in many fluid systems. Secondly, for translating obstacles, convective effects always dominate over viscous ones when one considers the flow sufficiently far from the obstacle \cite{proudman1957expansions,vandyke1975perturbation,veysey2007simple}. In two dimensions, this is what causes the Stokes paradox, thus preventing one from finding a Stokes solution \cite{lamb1932hydrodynamics}. In three dimensions, one can still obtain a Stokes solution but one is faced with the \emph{Whitehead paradox} \cite{an1889second,oseen1910uber}, which makes the low Reynolds number expansion a singular perturbation theory \cite{an1889second,vandyke1975perturbation,veysey2007simple}. Although resolving the Whitehead paradox leads one to find that Stokes solution accurately describes the flow at leading order in Reynolds number, this impediment motivates the need for understanding the role of convection in the case of odd viscous flow. Considering the role of convection for odd viscous flow is what is done in this work. Since the Whitehead paradox is related by the competition between convection and viscosity, complications arise when odd viscosity is taken into account, as now there is a competition between three distinct contributions. This complication is avoided because, similar to many Stokes flow computations \cite{olveira,khain2023trading,Khain_2022}, we solve the Navier-Stokes equations for low Reynolds numbers but also for small odd viscosity, so that at leading order the Whitehead paradox is resolved with the asymptotic matching procedure corresponding to purely shear viscous fluids. This solution can be corrected to account for odd viscous effects. 
\newline 
The structure of this work is as follows. In Sec.~\ref{sec:modeldescription} we describe the general three-dimensional odd viscous model that is being considered. In Sec.~\ref{sec:oseenliftdrag}, we study flow past a translating sphere at low Reynolds numbers. Specifically, we compute the \emph{odd oseenlet}, which serves as the first outer solution in an expansion based on asymptotic matching. In Sec.~\ref{sec:firstinnersolution}, we obtain the first inner solution exactly for the longitudinal case where \emph{axial symmetry} is upheld, and for the transverse case, where axial symmetry is broken, we use the Lorentz reciprocal theorem to calculate the first inertial correction to Stokes lift for the odd viscosity described in Sec.~\ref{sec:modeldescription}.

\section{Odd viscous Navier-Stokes equation}
\label{sec:modeldescription}
We consider an incompressible fluid system with a free-stream velocity $U_i$, a shear viscosity $\eta_s$, a constant density $\rho_0$ and a stationary and rigid no-slip sphere with radius $a$. We use these to non-dimensionalize the coordinate $x_i$, fluid velocity $u_i$ and stress tensor $\sigma_{ij}$. In addition, there is a single three-dimensional odd viscosity $\eta_o$ which is assumed to be small compared to $\eta_s$. We start from the steady, incompressible Navier-Stokes equation
\begin{subequations}
\label{eq:fullequationnavierstoke}
    \begin{align}
 \nabla_i u_i   & =0 ~~ ,  \\ 
  \text{Re} \,   u_j \nabla_j u_i   -    \nabla_j \sigma_{ij}    &  = 0   ~~  , 
\end{align}
\end{subequations}
where $\text{Re}$ is the Reynolds number, given by $\text{Re} = \rho_0 |U| a / \eta_s$. In three dimensions, for there to be odd viscosity, anisotropy is required. This anisotropy gives rise to multiple odd and even coefficients \cite{Khain_2022,khain2023trading}. To keep the computations tractable at nonzero Reynolds numbers, we consider only the simplest possible three-dimensional odd viscosity $\eta_o$ \cite{olveira,PhysRevLett.132.218303}, namely the odd viscosity that arises upon coarse-graining a system of particles with an intrinsic rotation whose rotation axis is pointed in direction $\ell_{i}$ \cite{markovich2021odd,markovich2022non}. Without loss of generality, we take this vector to be pointed in the $z$-direction, i.e. $\ell_i = \delta^z_i$. For such an odd viscosity and the isotropic shear viscosity, the stress $\sigma_{ij}$ constitutes of
\begin{align} \label{eq:constitutiveequation}
    \sigma_{ij} =  - p \delta_{ij}  + 2  \nabla_{(i} u_{j)}  + 2 \gamma_{\text{o}} (\delta_{i k} \varepsilon_{j l }  +  \varepsilon_{i k }  \delta_{j l }  )  \nabla_{(k} u_{l)} ~~  , 
\end{align}
where $p$ is the dimensionless pressure, $\gamma_{\text{o}} = \eta_o / \eta_s$. $\varepsilon_{i j } $ is the two-dimensional anisotropic Levi-Civita tensor, i.e. $\varepsilon_{i j }  = \epsilon_{i j  k } \ell_k  $, with $\epsilon_{i jk }$ being the three-dimensional Levi-Civita tensor. Plugging Eq.~\eqref{eq:constitutiveequation} into Eq.~\eqref{eq:fullequationnavierstoke} yields \cite{olveira,khain2023trading}
\begin{subequations}
\label{eq:fullequationnavierstoke1}
    \begin{align}
 \nabla_i u_i   & =0 ~~ ,  \\  \label{eq:differentialequation}
  \text{Re}\,    u_j \nabla_j u_i -   (   \delta_{ij}   +  \gamma_{\text{o}} \varepsilon_{ij}  ) \Delta u_j    +  \nabla_i p_{\text{m}}    &  = 0   ~~   , 
\end{align}
\end{subequations}
where $p_{\text{m}}$ we introduced the modified pressure \cite{ganeshan2017odd} given by $p_{\text{m}} = p  - \gamma_{\text{o}} \varepsilon_{ij} \nabla_i u_j $ and $\Delta$ is the Laplacian operator. Because we exclusively consider no-slip boundary conditions which do not depend on stress, it is more convenient to solve the fluid profile for the modified pressure rather than for the actual pressure. 

\section{Odd oseenlet}
\label{sec:oseenliftdrag}
We now consider the effect of a non-vanishing Reynolds number on the flow past an obstacle, which in the lab frame can be considered translating. We work in the co-moving frame, so that $U_i = U^{\infty}_i$, where $U^{\infty}_i$ is the far field fluid velocity as seen from the spherical obstacle. When considering low Reynolds number flow past a sphere, one runs into the Whitehead paradox \cite{an1889second,veysey2007simple,lamb1932hydrodynamics,vandyke1975perturbation}, which is the finding that is not possible to obtain a low Reynolds number correction to Stokes flow when one assumes the convective term $\propto\text{Re}$ in Eq.~\eqref{eq:differentialequation} to be globally negligible at leading order. To resolve this, we use \emph{asymptotic matching} \cite{proudman1957expansions,chester_breach_proudman_1969,kaplun1957low} which accounts for the fact that the convective term always dominates over viscous contributions far away from the object that causes the disturbance by finding inner and outer solutions to distinct expansions of Eq.~\eqref{eq:fullequationnavierstoke1}, which are matched to each other at the boundary of their respective regimes of validity.

\subsection{Zeroth inner solution}
For the inner solutions, we expand our fluid velocity as
\begin{subequations} \label{eq:innerexpansion}
    \begin{align} 
    u_i  & =   u_i^{(0)}  + \text{Re}\,   u_i^{(1)} + \mathcal{O}( \text{Re}^2 ) ~~  ,  \\ 
    p_{\text{m}} & =  p_{\text{m}}^{(0)}  + \text{Re}\,  p_{\text{m}}^{(1)} + \mathcal{O}(  \text{Re}^2 )~~  . 
 \end{align}
\end{subequations}
Eq.~\eqref{eq:innerexpansion} means that at leading order, Eq.~\eqref{eq:fullequationnavierstoke} is given by
 \begin{subequations}  \label{eq:fullequation12122}
\begin{align} 
\nabla_i u^{(0)}_i   & =0 ~~ ,  \\ 
\label{eq:ffeuiheiuh}
  (  \delta_{ij}   +  \gamma_{\text{o}}  \varepsilon_{ij}  ) \Delta  u^{(0)}_j   -  \nabla_i p^{(0)}_{\text{m}}  & = 0   ~~ , \end{align}
\end{subequations}
Working in the co-moving frame with the sphere's center as the origin, the boundary conditions we impose are given by
\begin{align} \label{eq:boundarycondition}
    \lim_{r \rightarrow \infty} u_{i} = e_i ~~ , ~~     u_{i}|_{r = 1 } = 0  ~~  , 
\end{align}
where $r = \sqrt{x^2 + y^2 + z^2 }$ and $e_i = U^{\infty}_i / |U^{\infty}| $. Eq.~\eqref{eq:fullequation12122} has been solved for the boundary conditions of Eq.~\eqref{eq:boundarycondition} in Refs.~\cite{hosaka2023lorentz,PhysRevLett.132.218303,olveira} by going to Fourier space. We now show a distinct derivation which does not require going to Fourier space because we wish to use the same approach to derive the odd oseenlet in App.~\ref{sec:oddoseen}. To understand how \eqref{eq:fullequation12122} is solved for flow around a sphere, we first note that far away from the sphere, it can be treated as if the total Stokes force $  \mathcal{F}^{(0)}_i$ were localized at a singular point \cite{Kimkarrila2013-op}. We can thus rewrite Eq.~\eqref{eq:fullequation12122} as
\begin{subequations}  \label{eq:fullequationnaoke1}
\begin{align} 
\nabla_i u^{\prime (0)}_i   & =0 ~~ ,  \\ 
\label{eq:ffeuiheiuhiufee}
  (  \delta_{ij}   +  \gamma_{\text{o}}  \varepsilon_{ij}  ) \Delta  u^{\prime (0)}_j   -  \nabla_i p^{\prime (0)}_{\text{m}}  & =    \mathcal{F}^{(0)}_i \delta    ~~ , \end{align}
\end{subequations}
where $\delta$ is the Dirac delta function located at $x_i=0$ and and $  \mathcal{F}^{(0)}$ is the dimensionless total force on the sphere. We furthermore introduced primed point-force fields, which are related to the original fields as 
\begin{align}
    u^{\prime (0)}_i   =  u^{ (0)}_i  + \mathcal{O} (r^{-2}) ~~ , ~~    p_{\text{m}}^{\prime (0)}   =  p_{\text{m}}^{ (0)}  + \mathcal{O} (r^{-3}) ~~ ,   
\end{align}
We first solve for pressure by taking the divergence of Eq.~\eqref{eq:fullequationnavierstoke1}, this yields
\begin{align}
      p^{\prime (0)}_{\text{m}}    =    \mathcal{F}^{(0)}_i  
 \nabla_i \frac{1}{4 \pi   r} +   \gamma_o \varepsilon_{ij}  \nabla_i  u^{\prime (0)}_j ~~ ,   \end{align}
 where we used that $ \delta =   -   \Delta  \frac{1}{4 \pi  r }  $. Plugging this back into Eq.~\eqref{eq:ffeuiheiuhiufee}, we obtain
 \begin{align}  \label{eq:oseenequation2122322}
  \begin{split}
    &  \left[  \delta_{ij}    \Delta  + \gamma_{\text{o}}  ( \varepsilon_{ij}  \Delta  -  \varepsilon_{k j }   \nabla_i   \nabla_k )  \right]    u^{\prime (0)}_j  \\  &  +     \frac{  \mathcal{F}^{(0)}_j}{4 \pi }  \left(  \delta_{ij}   \Delta   -   \nabla_i
  \nabla_j    \right) \frac{1}{ r }  =0     ~~ .
   \end{split}  
\end{align}
We solve for $ u^{\prime (0)}_j$ perturbatively in $\gamma_{\text{o}}$ \cite{Khain_2022,olveira,hosaka2023lorentz}. For this, we define for a general function $f$
    \begin{align}
f  & =      f^{(\text{s})}  +   \gamma_{\text{o}}   f^{(\text{o})}  + \mathcal{O} (\gamma^2_o) ~~ ,     
\end{align}
At zeroth order in $\gamma_o$, we can then find the point-force solution given by \cite{Kimkarrila2013-op}
\begin{subequations} \label{eq:leadingordersolutions}
    \begin{align}
  u^{\prime (0,s)}_i   & =  e_i -    \frac{1}{8 \pi }    G^{(0,\text{s})}_{ij}   \mathcal{F}^{(0,\text{s})}_j  ~~    , 
\end{align}
\end{subequations}
where $G^{(0,\text{s})}_{ij} $ is the Oseen tensor, which can be expressed as
\begin{align}
    G^{(0,\text{s})}_{ij}   = \left(  \delta_{ij} \Delta   - \nabla_i
 \nabla_j    \right)  r ~~ . 
\end{align}
To obtain $ u^{\prime (0,\text{o})}_i$, we plug back Eq.~\eqref{eq:leadingordersolutions} into Eq.~\eqref{eq:oseenequation2122322}, yielding
 \begin{align}  \label{eq:oseenequa322}
  \begin{split}
    &        \Delta   u^{\prime (0,\text{o})}_i   +   ( \varepsilon_{ij}  \Delta  \rl{-}  \varepsilon_{k j }   \nabla_i   \nabla_k )    u^{\prime (0,\text{s})}_j  \\  &  +     \frac{  \mathcal{F}^{(0,\text{o})}_j}{4 \pi }  \left(  \delta_{ij}   \Delta   -   \nabla_i
  \nabla_j    \right) \frac{1}{ r }  =0     ~~ .
   \end{split}  
\end{align}
Eq.~\eqref{eq:oseenequa322} is solved by
    \begin{align} \label{eq:stokesle}
     u^{\prime (0,\text{o})}_i   &  =  -  \frac{1}{8 \pi }  \left(   G^{(0,\text{o})}_{ij}   \mathcal{F}^{(0,\text{s})}_j +   G^{(0,\text{s})}_{ij}   \mathcal{F}^{(0,\text{o})}_j   \right)   ~~   , 
\end{align} 
where $G^{(0,\text{o})}_{ij}$ is the odd Oseen tensor \cite{olveira} given by
    \begin{align}
  G^{(0.\text{o})}_{ij} &   =    -   \left(      \varepsilon_{ij} \delta_{kl}
  + 2 \delta_{l [ i } \varepsilon_{j] k}   
  \right) \nabla_k \nabla_l  r ~~ . 
\end{align}
Having found the point-force solution which provides the leading order fluid profile contributions away from the sphere, let us now go back to the question of how the Stokes equation is solved without assuming that one is far away from the sphere. Because of the symmetric shape of the sphere, the leading order correction to the point-force induced flow is given by the Laplacian of the point-force solution. Specifically, imposing the boundary condition $  u_{i}|_{r = 1 } = 0$ leads one to find that the odd viscous flow past a sphere is given by \cite{hosaka2023lorentz}
\begin{align} \label{eq:spheresolution}
\begin{split} 
             u^{(0)}_i      =  & e_i -  \frac{1}{8 \pi } \left(1  + \frac{1}{6} \Delta^2  \right)   \left( G^{(0,\text{s})}_{ij}   + \gamma_o  G^{(0,\text{o})}_{ij}  \right)   \mathcal{F}^{(0)}_j   \\  & 
             + \mathcal{O} (\gamma_o^2)  ~~  , 
             \end{split}
\end{align}
where the corresponding total Stokes force is required to be \cite{hosaka2023lorentz,PhysRevLett.132.218303,khain2023trading}
\begin{align}
      \mathcal{F}^{(0)}_i = \left(   C^{(0)}_D \delta_{ij}  + C^{(0)}_L \varepsilon_{ij} \right) e_j   ~~ , 
\end{align}
with drag and lift coefficients
\begin{align} \label{eq:coefficients}
    C_D^{(0)}  =   6 \pi ~~ , ~~ C_L^{(0)}  =    3 \pi  \gamma_{\text{o}}  ~~ . 
\end{align}
Restoring the dimensionality and defining the total force $F_i$, this means
\begin{align}
    F_i = 6 \pi a  \eta_s    \left(  \delta_{ij}  +  \frac{1}{2} \gamma_o    \varepsilon_{ij} + \mathcal{O} ( \gamma_o^2, \text{Re})  \right) U^{\infty}_j   ~~ , 
\end{align}

\subsection{First outer solution}
Having found the zeroth inner solution, we connect this solution to the outer region to obtain the first outer solution. The boundary outer region is characterized by $ \text{Re} \,  r    =  \mathcal{O } (1)$, so that in order to consistently match small Reynolds number corrections, we work with the Oseen coordinate $ \tilde  r   = \text{Re} \,  r $. For the outer solutions, we expand the fields as
\begin{subequations} 
    \begin{align}
    \tilde u_i  &  = e_i   + \text{Re} \,  \tilde u^{(1)}_i + \mathcal{O}( \text{Re}^2 )  ~~ ,  \\  
        \tilde p_{\text{m}}  & =   Re^2  \,   \tilde p^{(1)}_{\text{m}} + \mathcal{O}( \text{Re}^3 )  ~~ . 
\end{align}
\end{subequations}
The equation that dictates the first outer solution is the Oseen equation \cite{lamb1932hydrodynamics,oseen1910uber}, to wit
\begin{subequations}  \label{eq:fullequation1233214}
\begin{align}
\tilde \nabla_i \tilde  u^{(1)}_i   & =0 ~~ ,  \\ 
  (   \delta_{ij}   +  \gamma_{\text{o}} \varepsilon_{ij}  ) \tilde \Delta \tilde u^{(1)}_j   -   \tilde \nabla_i p^{(1)}_{\text{m}}  & =       e_j \tilde   \nabla_j \tilde  u^{(1)}_i    ~~ , 
\end{align}
\end{subequations}
Asymptotic matching dictates that we must connect $ u^{(1)}_i$ to the zeroth inner solution. We find the corresponding inner boundary condition to be given by
\begin{subequations}  \label{eq:outerboundarycond}
    \begin{align}
    \begin{split}
        \lim_{\tilde r \rightarrow 0 }  \tilde  u^{(1)}_i     = &     - \frac{1   }{8 \pi } \left( \tilde  G^{(0,\text{s})}_{ij}   + \gamma_o  \tilde G^{(0,\text{o})}_{ij}  \right)   \mathcal{F}^{(0)}_j   \\ &  +  \mathcal{O}( \gamma_{\text{o}}^2)  
                  \end{split}
  \\ \begin{split}
         \lim_{\tilde r \rightarrow 0 }          \tilde  p^{(1)}   =  &    \frac{1}{4 \pi }      \mathcal{F}^{(0)}_i  
 \tilde \nabla_i \frac{1}{\tilde r}    - \frac{\gamma_o}{8 \pi } \varepsilon_{ij}  \tilde \nabla_i \tilde G^{(0,\text{s})}_{j l }   \mathcal{F}^{(0)}_l   \\  &  +    \mathcal{O}( \gamma_{\text{o}}^2)~~ , 
   \end{split}
\end{align}
\end{subequations}
where 
    \begin{subequations}
    \begin{align}
\tilde   G^{(0,\text{s})}_{ij}   & = \left(  \delta_{ij} \tilde  \Delta   - \tilde  \nabla_i \tilde 
 \nabla_j    \right)  \tilde  r  \\ 
\tilde   G^{(0.\text{o})}_{ij} &   =    -   \left(      \varepsilon_{ij} \delta_{kl}
  + 2 \delta_{l [ i } \varepsilon_{j] k}   
  \right) \tilde \nabla_k \tilde \nabla_l  \tilde r ~~ . 
\end{align}
\end{subequations}
Eq.~\eqref{eq:outerboundarycond} together with the outer boundary condition and Eq.~\eqref{eq:fullequation1233214} force $   \tilde  u^{(1)}_i$ to be given by
\begin{subequations} \label{eq:firstoseensolutionfull}
    \begin{align} \label{eq:firstoseensolution}
         \tilde  u^{(1)}_i   & =   -   \frac{1}{8 \pi } \left(\tilde  G^{(1,\text{s})}_{ij}   + \gamma_{\text{o}}  \tilde G^{(1,\text{o})}_{ij}  \right)   \mathcal{F}^{(0)}_j   +  \mathcal{O}( \gamma_{\text{o}}^2) ~~  , \\ 
      p^{(1)}_m    & =  \frac{1}{4 \pi }      \mathcal{F}^{(0)}_i  
\tilde \nabla_i \frac{1}{\tilde r}    - \frac{\gamma_o}{8 \pi } \varepsilon_{ij}  \tilde \nabla_i \tilde G^{(1,\text{s})}_{j l }   \mathcal{F}^{(0) }_l        +  \mathcal{O}( \gamma_{\text{o}}^2) ~~  , 
\end{align}
\end{subequations}
where $\tilde G^{(1,\text{s})}_{ij}$ represents the even oseenlet of the form \cite{Pozrikidis1996-dq}
\begin{align}
\tilde    G^{(1,\text{s})}_{ij}  =   2   \left(  \delta_{ij} \tilde  \Delta   - 
 \tilde \nabla_j \tilde  \nabla_i  \right)    \int_0^{\tilde \zeta }  \frac{1 - \exp(-  \xi)}{\xi} d \xi    ~~ , 
\end{align}
with $\tilde \zeta =  \frac{1}{2}(  \tilde r  - \tilde x_i e_i )$. $\tilde G^{(1,\text{o})}_{ij}$ is the odd oseenlet, which is given by (see App.~\ref{sec:oddoseen} for a derivation)
\begin{align}
   \tilde  G^{(1,\text{o})}_{ij}  =  2  \left(      \varepsilon_{ij} \delta_{kl}
  + 2 \delta_{l [ i } \varepsilon_{j] k}   
  \right)  \tilde \nabla_k  \tilde \nabla_l   \exp(- \tilde \zeta   )     ~~ . 
\end{align}

\section{First inner solution}
\label{sec:firstinnersolution}
Having found the first outer solution, we use asymptotic matching to obtain the outer boundary condition for $u^{(1)}_i$. Asymptotic matching of the inner solution to the outer solution at $\mathcal{O} (\text{Re})$ leads to the outer boundary condition 
\begin{subequations}  \label{eq:firstinnerconstraintfull}
    \begin{align} \label{eq:firstinnerconstraint}
 \lim_{r \rightarrow \infty }  u^{(1)}_i   &    =   \frac{1}{ 8 \pi    }   \mathcal{F}^{(0)}_j  \left[     D^{(s)}_{ij}
  +  \gamma_{\text{o}}  D^{(o)}_{ij}     \right]   \zeta^2     +  
 \mathcal{O} (\gamma_{\text{o}}^2 )   ~~ ,  \\  \label{eq:firstinnerconstraint1}
 \lim_{r \rightarrow \infty }  p^{(1)}_{\text{m}}   &  = -   \frac{\gamma_o }{ 16 \pi    } 
 \varepsilon_{ij}      \mathcal{F}^{(0)}_i  \nabla_j \Delta       \zeta^2 +  
 \mathcal{O} (\gamma_{\text{o}}^2 )  
      ~~  , 
 \end{align}
 \end{subequations}
 where $\zeta =  \frac{1}{2}( r  -x_i e_i )$ and we used the definitions
 \begin{subequations}
      \begin{align}
     D^{(\text{s})}_{ij}   & =  \frac{1}{2} ( \delta_{ij} \Delta   - 
 \nabla_j  \nabla_i )    ~~ , \\ 
    D^{(\text{o})}_{ij}    &  =  - \left(      \varepsilon_{ij} \delta_{kl}
  + 2 \delta_{l [ i } \varepsilon_{j] k}   
  \right)   \nabla_k   \nabla_l  ~~ . 
 \end{align}
  \end{subequations}
Expanding Eq.~\eqref{eq:fullequationnavierstoke} up $\mathcal{O}(\text{Re})$, one finds that the equations that the first inner solution must satisfy a differential equation given by
\begin{subequations}  \label{eq:fullequation124}
\begin{align}
 \nabla_i   u^{(1)}_i   & =0 ~~ ,  \\  \label{eq:innerequation}
  (  \delta_{ij}   +  \gamma_{\text{o}}  \varepsilon_{ij}  ) \Delta  u^{(1)}_j   -  \nabla_i p^{(1)}_{\text{m}} & =  u^{(0)}_j   \nabla_j    u^{(0)}_i    ~~ .  
\end{align}
\end{subequations}
Until now, we made no assumption about the \emph{direction} of the free-stream velocity with respect to $\ell_i$. In order to compute the first inner solution, it is important that we do so. Let us first consider the longitudinal case where $e_i = \ell_i$. The longitudinal case is simpler than the transverse case as we retain axial symmetry. For the case with transverse free-stream velocity, we refrain from computing the first inner solution. We instead compute the lift force, which is exists only for the transverse case, directly using the Lorentz reciprocal theorem.

\begin{figure}[t]
    \centering \includegraphics[width=0.7\linewidth]{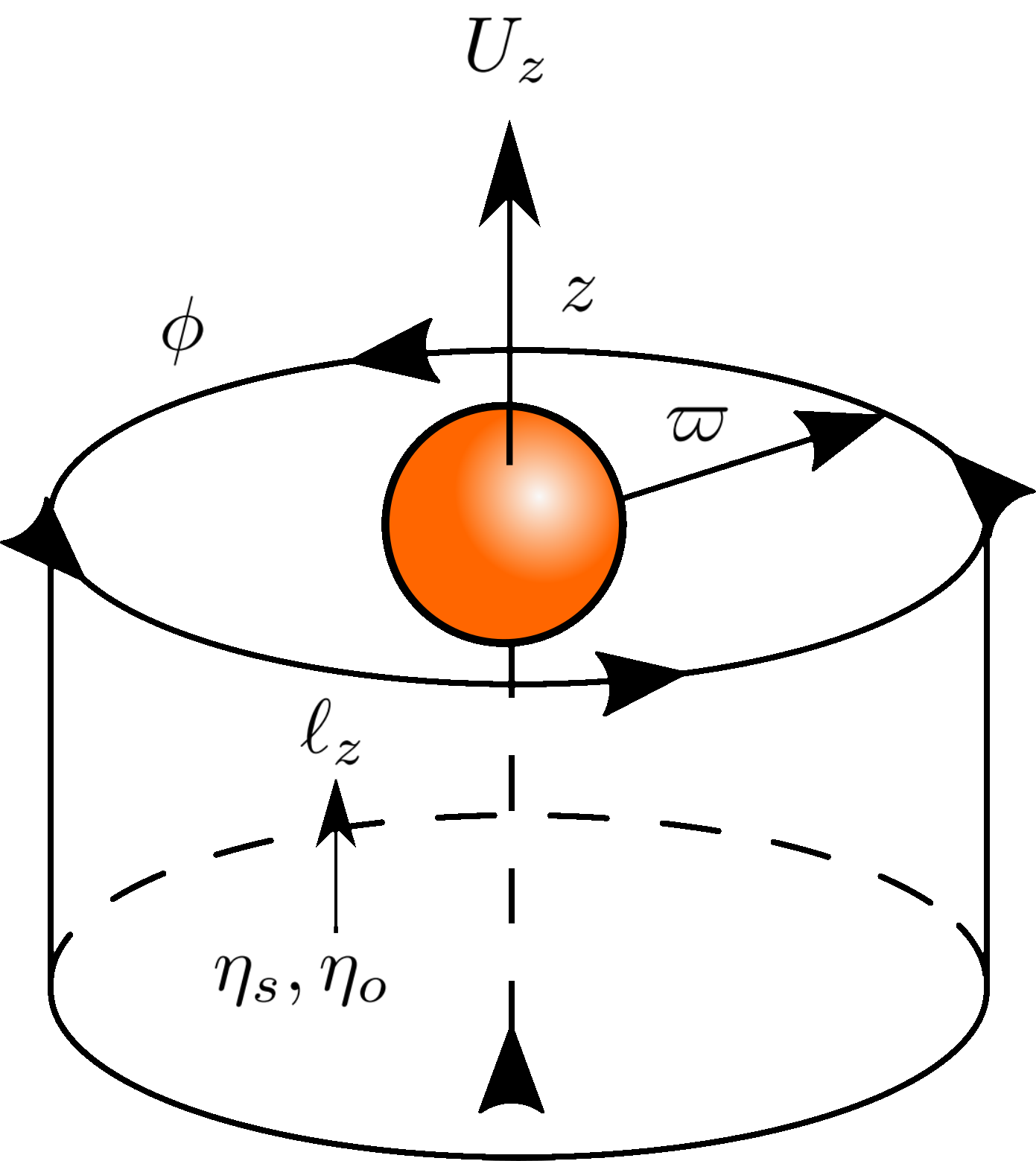}
\caption{Graphical representation of cylindrical coordinates $\varpi,\phi,z$ for a sphere translating with velocity $U_z$ immersed in a fluid with viscosities $\eta_s$, $\eta_o$. The anisotropy vector $\ell_i$ of the odd viscous fluid is also pointed in the $z$-direction.}    \label{fig:enter-label2981}
\end{figure}
\subsection{Longitudinal free-stream velocity}
For the case $e_i = \ell_i$, let us first solve Eq.~\eqref{eq:fullequation124} at zeroth order in $\gamma_o$. For this, we use the axial symmetry of the longitudinal free-stream velocity to take the stream function ansatz \cite{happel1983low}
\begin{align}
    u^{(1,\text{s})}_{\varpi }   =  -   \frac{ \partial_{z }\psi^{(1,\text{s})}}{\varpi }    ~~ , ~~    u^{(1,\text{s})}_{\phi }  =0  ~~,  ~   u^{(1,\text{s})}_{z}   =    \frac{ \partial_\varpi \psi^{(1,\text{s})}}{ \varpi  }  , 
\end{align} 
where $\varpi,\phi$ and $z$ are cylindrical coordinates as visualized in Fig.~\ref{fig:enter-label2981}. The solution that Eq.~\eqref{eq:fullequation124} and satisfies the inner and outer boundary is given by \cite{proudman1957expansions,vandyke1975perturbation}
\begin{align}
\begin{split}
        \psi^{(1,\text{s})}  =  &  \frac{3}{32} \varpi ^2 \left(\frac{1}{r^3}-\frac{3}{r}+2\right) \\  & -\frac{3}{32} \varpi ^2 z\left(\frac{1}{r^5}-\frac{1}{r^4}+\frac{1}{r^3}-\frac{3}{r^2}+\frac{2}{r}\right)  ~~, 
        \end{split}
\end{align}
with $r =\sqrt{\varpi^2 + z^2 }$. Now we move to $ u^{(1,\text{o})}_{i} $. Firstly, we define $ u^{(1,\text{o})}_{i}  =  u^{\prime (1,\text{o})}_{i} +  u^{\prime \prime (1,\text{o})}_{i} $, with 
\begin{align}
     u^{\prime (1,\text{o})}_{i}   = 0  ~~ , ~~    u^{ \prime (1,\text{o})}_{\phi }  =  -   \frac{ \partial_{z }\psi^{(1,\text{s})}}{\varpi }    ~~,  ~   u^{ 
 \prime (1,\text{o})}_{z }   =   0  ~~ , 
\end{align}
so that at $\mathcal{O}(\gamma_o)$ Eq.~\eqref{eq:innerequation} reduces to
\begin{align}
\label{eq:innerequation1}
 \Delta  u^{\prime \prime (1,\text{o})}_i    -  \nabla_i p^{(1,\text{o})}_{\text{m}} & =  u^{(0,\text{s})}_j   \nabla_j    u^{(0,\text{o})}_i + u^{(0,\text{o})}_j   \nabla_j    u^{(0,\text{s})}_i    ~~ .
\end{align}
The right hand side of Eq.~\eqref{eq:innerequation1} only has a non-vanishing $\phi$-component and thus $u^{(1,\text{o})}_i$  also only has a $\phi$-component which is automatically incompressible due to axial symmetry, i.e. $\nabla_i u^{(1,\text{o})}_i = \varpi^{-1} \partial_{\phi } u^{(1,\text{o})}_{\phi} =0 $. Similarly, axial symmetry forces $p^{(1,\text{o})}$ to be a constant. Lastly, noting that $u^{(0,s)}_{\phi} =0 $, Eq.~\eqref{eq:innerequation1} thus reduces to
\begin{align}
\label{eq:innerequation2}
    \Delta u^{\prime \prime (1,\text{o})}_{\phi}   & =  
   u^{(0,\text{s})}_j   \nabla_j    u^{(0,\text{o})}_{\phi }     ~~ .  
\end{align}
Before we obtain $u^{\prime \prime  (1,\text{o})}_{\phi}$, let us consider the outer boundary condition of Eq.~\eqref{eq:firstinnerconstraint}. Since it holds that
\begin{align}
    \lim_{r \rightarrow \infty } (u^{(1,\text{o})}_{\phi} - u^{\prime (1,\text{o})}_{\phi}  )  = \frac{3  \varpi ^3}{16  r^3 } ~~ ,  
\end{align}
we must have
\begin{align} \label{eq:secondboundarycondition}
    \lim_{r \rightarrow \infty } u^{\prime \prime  (1,\text{o})}_{\phi} = \frac{3  \varpi ^3}{16  r^3 }  ~~ .   
\end{align}
To solve Eq.~\eqref{eq:innerequation2} with the given boundary conditions, we define $u^{\prime \prime (1,\text{o})}_{\phi} = \Phi_P + \Phi_H $. $\Phi_P$ must obey the equation
\begin{align}
\label{eq:innerequation21}
\mathcal{D} \Phi_P   & =  
   u^{(0,\text{s})}_j   \nabla_j    u^{(0,\text{o})}_{\phi }     ~~ , 
\end{align}
with $\mathcal{D}   =  \partial_z^2  - \frac{1}{\varpi^2 }  + \frac{1}{\varpi} \partial_{\varpi } + \partial_{\varpi}^2  $. The solution to Eq.~\eqref{eq:innerequation21} is given by
    \begin{align}\label{eq:particular}
    \begin{split}
            \Phi_P  =   & \frac{1}{32} \varpi \left(-\frac{1}{r^6}-\frac{12}{r^4}\right)  \\  & 
            +\frac{1}{32}  \varpi ^3 \left(\frac{3}{4 r^8}+\frac{18}{r^6}-\frac{12}{r^5}-\frac{9}{2 r^4}+\frac{6}{r^3}\right)  ~~ .     \end{split}
\end{align}    
Having obtained the particular solution, we look for the homogeneous solution. Note that 
\begin{align} \label{eq:plimit}
    \lim_{r \rightarrow \infty }   \Phi_P  =  \frac{3  \varpi ^3}{16  r^3 }  ~~, 
\end{align}
so that we require
\begin{align} \label{eq:Honstraint}
      \lim_{r \rightarrow \infty }   \Phi_H  =0  ~~ . 
\end{align}
$ \Phi_H $ obeys the equation
 \begin{align} \label{eq:differentialequation12}
  \mathcal{D}  \Phi_H  =0 ~~. 
 \end{align}
  Eq.~\eqref{eq:differentialequation12} is solved by
 \begin{align}  \label{eq:Hsolution}
\Phi_H  = A_1  \frac{ \varpi }{r^3}+A_2 \left(\frac{\varpi }{r^5 }-\frac{5 \varpi ^3}{4 r^7 }\right)~~, 
\end{align}  
To satisfy the inner no-slip boundary condition of Eq.~\eqref{eq:boundarycondition}, we must take $A_1 = \frac{1}{5} $ and $A_2 = \frac{33}{160}$. Having obtained a complete solution up to first order in odd viscosity and Reynolds number, we then compute the force on the sphere corresponding to this solution. As follows from axial symmetry, there is no lift force or any other linear odd correction to the force, to wit
\begin{align}
F_i = 6 \pi a  \eta_s | U^{\infty}|    \left( 1  +    \frac{3}{8} \text{Re}  + \mathcal{O} ( \gamma_o^2 , \text{Re}^2 )  \right) \ell_i   ~~ . 
\end{align}
However, we do find odd viscous effects through the following dimensionless \emph{torque}
\begin{align} \label{eq:totaltorque}
    \mathcal{T}^{(1)}_{k}   =  \int_{B}   \epsilon_{ij k } x_i  \sigma^{(1)}_{jl} d S_l     = \frac{2 \pi}{5  } \gamma_o   +   \mathcal{O}(\gamma_o^2 )  ~~ . \end{align}
 Restoring dimensions, the total torque is given by
 \begin{align}
   T_i    = \frac{2 \pi }{5}    a^2  \eta_s  | U^{\infty}|  \left(  \gamma_o  \text{Re}      + \mathcal{O} ( \gamma_o^2 , \text{Re}^2 )  \right) \ell_i  ~~ . 
 \end{align}
 We thus find that convection can allow for a sphere to feel torque when \emph{moving through} an odd viscous fluid in an axially symmetry way. Reflection symmetry prevents such a torque from arising for Stokes flow past spheres \cite{khain2023trading}.

\subsection{Transverse free-stream velocity}
For the case where the free-stream velocity is transverse to $\ell_i$, to explicitly solve Eq.~\eqref{eq:fullequation124} in a way that satisfies the inner and outer boundary conditions is challenging, in part because this forbids usage of stream functions. Fortunately, it turns out that we need not know more about the first inner solution than Eqs.~\eqref{eq:firstinnerconstraintfull} and \eqref{eq:fullequation124} in order to obtain the first convective correction to drag and lift force on a translating sphere. This because we can use the Lorentz reciprocal theorem to directly compute these forces instead of obtaining them from the force density corresponding to the first inner solution \cite{masoud2019reciprocal,Brenner_Cox_1963}. The Lorentz reciprocal theorem connects two fluid systems, with the first fluid system being the one we wish to better understand and an \emph{auxiliary fluid system} that is used to accomplish this by means of the Lorentz reciprocal theorem. The first fluid system is one corresponding to the first inner solution represented by fluid velocity $u^{(1)}_i$. The auxiliary fluid system labelled by $*$ is identical to the first fluid system except for the following three things:
\begin{enumerate}
\item We consider for the auxiliary fluid system a dimensionless free-stream velocity $e^{*}_i$ which satisfies $(e^{*}_i)^2=1$ and $e^{*}_i \ell_i=0$, but is not necessarily parallel to $e_i$, as a parallel $e^{*}_i$ would not allow one to extract lift force using the Lorentz reciprocal theorem \cite{lier2024slipinduced}. 
    \item For the auxiliary fluid system, we consider Stokes flow, i.e. flow which corresponds to the zeroth inner solution represented by fluid velocity $u^{* (0)}_i$.
\item In order for the Lorentz reciprocal theorem to hold for odd fluids, we require that odd viscosity of the second fluid system is given by $\gamma^{*}_{\text{o}} =  - \gamma_{\text{o}}$ \cite{hosaka2023lorentz}. 
\end{enumerate}
\begin{figure}[t]
    \centering \includegraphics[width=1\linewidth]{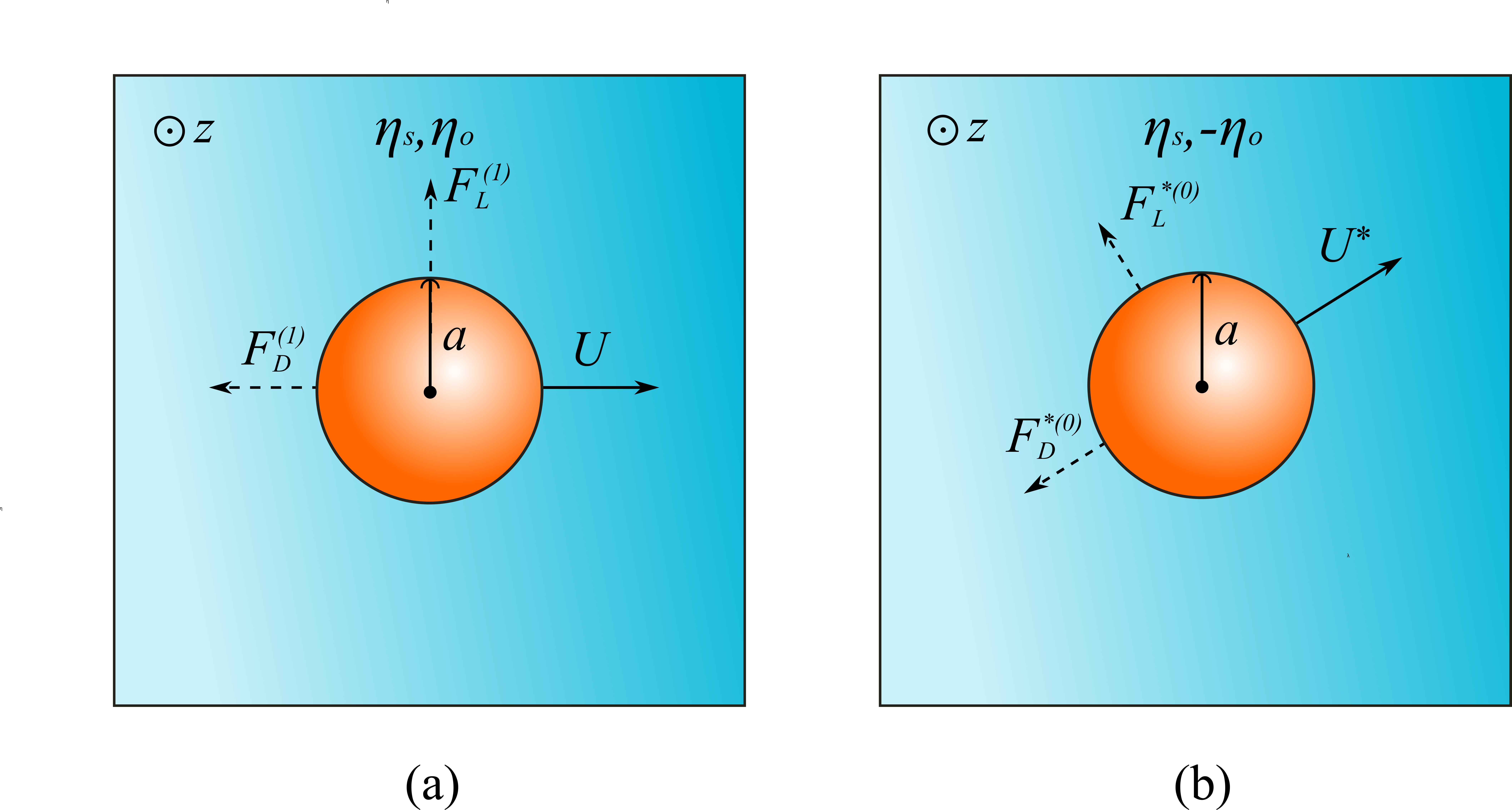}
\caption{Graphical representation of the two fluid systems that are considered when using the Lorentz reciprocal theorem. These are (a) the first fluid system which we wish to understand better and (b) the auxiliary fluid system which is used to gain understanding of the first one by means of the Lorentz reciprocal theorem.}    \label{fig:LRT}
\end{figure}
These differences are represented in Fig.~\ref{fig:LRT}. Starting from Eq.~\eqref{eq:fullequation124}, the Lorentz reciprocal theorem gives the relation \cite{Brenner_Cox_1963,masoud2019reciprocal}
 \begin{align}  \label{eq:reciprocalequations}
    \begin{split}
         & \int_{S_{\infty}}   \left(   u_j^{* (0)}  \sigma^{ (1)}_{ij}    -    u_j^{ (1)} \sigma^{* (0)}_{ij}   \right) d S_i   =     \int_{V_{\infty}}     u^{ (0)}_i  \nabla_i u^{ (0)}_j   u^{* (0)}_{j}  d V    ,  \end{split}
\end{align}
 where $S_{\infty }$ is the surface of an infinitely large sphere formed around the sphere with no-slip boundary conditions and $V_{\infty}$ is the volume enclosed by $S_{\infty}$ and surrounding $B$, the boundary of the no-slip sphere. To work out Eq.~\eqref{eq:reciprocalequations}, we introduce the definitions
\begin{subequations}
     \begin{align}  \label{eq:recip123}
      I_1   =   & \int_{S_{\infty}}      u_j^{* (0)}  \sigma^{ (1)}_{ij}   d S_i  ~~ ,  \\ 
          I_2   =   &   -  \int_{S_{\infty}}    u_j^{ (1)} \sigma^{* (0)}_{ij}  d S_i   ~~ , \\   \label{eq:thirdterm}
          I_3=  &   \int_{V_{\infty}}     u^{ (0)}_i  \nabla_i u^{ (0)}_j   u^{* (0)}_{j}  d V
\end{align}
\end{subequations}
`For the first two terms it holds that because the sphere with surface $S_{\infty}$ is infinitely large, any contribution to the integrand which is $\mathcal{O} (r^{-3})$ can be ignored. To take advantage of this fact, let us introduce for a general field $f$ the notation
\begin{align}
    f = \sum_{n=0} ( f  )_{-n} r^{-n}  ~~,
\end{align}
where $(f)_{-n}$ has no further $r$-dependence. We first consider $I_1$, which can be rewritten as
\begin{align}  
\begin{split}
    \label{eq:I1simplification}
      I_1   =   &    \int_{S_{\infty}}   (\sigma^{ (1)}_{ij})_{-1}     
  \left[  ( u^{* (0)}_j)_0 r  + ( u^{* (0)}_j)_{-1}     r^{-2}       \right]  d S_i  +  \\ 
   &  +    \int_{S_{\infty}}    (\sigma^{ (1)}_{ij})_{-2} ( u^{* (0)}_j)_{0}   r^{-2} d S_i  ~~   . 
 \end{split}
\end{align}
$  u^{* (0)}_j $ is fully even under $x_i \rightarrow -x_i $ and since Eq.~\eqref{eq:I1simplification} is a surface integral, $  (\sigma^{ (1)}_{ij})_{-1}     $ and $  (\sigma^{ (1)}_{ij})_{-2}     $ must be odd under $x_i \rightarrow -x_i $ in order for $I_1$ to be non-vanishing. $(\sigma^{ (1)}_{ij})_{-1}$ is fully determined by $(p_{m})_{-1}$ and $( u^{ (1) }_j  )_{0}$, which are given by Eq.~\eqref{eq:firstinnerconstraintfull}. Eq.~\eqref{eq:firstinnerconstraint1} tells us that $(p_{m})_{-1}$ is even under $x_i \rightarrow -x_i $. For the remaining part of the stress to give a contribution odd under $x_i \rightarrow -x_i $ requires $( u^{ (1) }_j  )_{0}$ to be even under $x_i \rightarrow -x_i $. From Eq.~\eqref{eq:firstinnerconstraint1} it follows that $( u^{ (1) }_j  )_{0}$ has a part that is even and odd under $x_i \rightarrow -x_i $, but the part that is even under $x_i \rightarrow -x_i $ is constant. We define this constant part to be $ J^{(1)}_{jk}   \mathcal{F}^{(0)}_k$. From Eq.~\eqref{eq:firstinnerconstraint1} we learn that $J^{(1)}_{jk} $ is given by
\begin{align} \label{eq:simpleidentity}
    J^{(1)}_{ij}  =  \frac{1}{32 \pi   } \left(3 \delta_{ij}  - e_i e_j -  2  \gamma_o  \varepsilon_{ij} \right)  ~~ . 
\end{align}
Because this term is constant, it vanishes when computing the corresponding viscous stress, making $(\sigma^{ (1)}_{ij})_{-1}$ uniformly even under $x_i \rightarrow -x_i $. Therefore, this contribution vanishes, and Eq.~\eqref{eq:I1simplification} simplifies to
\begin{align}  
\begin{split}
    \label{eq:I1simplification1}
      I_1   =  &  e^{*}_j  \int_{S_{\infty}}        \sigma^{ (1)}_{ij}  d S_i ~~   . 
 \end{split}
\end{align}
$  I_1$ can thus be related to the desired $\mathcal{F}^{(1)}_i$ whose expression is given by
\begin{align} 
    \mathcal{F}^{(1)}_i =   \int_{B } \sigma^{ (1)}_{ij}   d S_j ~~ . 
\end{align}
Specifically, from Eq.~\eqref{eq:innerequation} it follows that
\begin{align} \label{eq:gaussianrelation}
    \int_{B + S_{\infty}} \left(\sigma^{ (1)}_{ij}  - u^{(0)}_j  u^{(0)}_j   \right) d S_j   =0  ~~ .  
\end{align}
Eq.~\eqref{eq:gaussianrelation} simplifies by noting that $u^{(0)}_j   =0  $ on $B$. Furthermore, since $u^{(0)}_j $ is fully even under $x_i \rightarrow - x_i $, we have
\begin{align}
  \int_{B + S_{\infty}}     u^{(0)}_j  u^{(0)}_j    d S_j   = 0  ~~ , 
\end{align}
leaving us with
\begin{align} \label{eq:forceequation1}
       \mathcal{F}^{(1)}_i =-    \int_{S_{\infty}} \sigma^{ (1)}_{ij}   d S_j    ~~ , 
\end{align}
Eq.~\eqref{eq:gaussianrelation} can then be used to obtain
\begin{align}
   I_1 =    - e^{*}_j       \mathcal{F}^{(1)}_j ~~ . 
   \end{align}
We then consider $I_2$, which can be rewritten as
\begin{align}
    I_2 = -  \int_{S_{\infty}}   r^{-2} ( u_j^{ (1)} )_0 
 ( \sigma^{* (0)}_{ij} )_{-2} d S_i   ~~  , 
\end{align}
because $( \sigma^{* (0)}_{ij} )_{-1} =0 $. $( \sigma^{* (0)}_{ij} )_{-2} $ is odd under $x_i \rightarrow - x_i $, so that only the even part of $(u^{(1)}_j)_0$ can contribute. Therefore, using Eq.~\eqref{eq:simpleidentity}, $I_2$ simplifies to
\begin{align} \label{eq:simpleI2}
    I_2 =  -  J^{(1)}_{jk}    \mathcal{F}^{(0)}_k  \int_{S_{\infty}}    
    \sigma^{* (0)}_{ij} d S_i   ~~  , 
\end{align}
Similar to Eq.~\eqref{eq:forceequation1}, it holds for the zeroth order Stokes force that 
\begin{align} \label{eq:relation}
      \mathcal{F}^{* (0)}_{i}    =  -  \int_{S_{\infty}} \sigma^{* (0)}_{ij}  d S_j   ~~ , 
\end{align}
so that Eq.~\eqref{eq:simpleI2} turns into
\begin{align} \label{eq:simplerI2}
    I_2 =  - J^{(1)}_{jk}   
     \mathcal{F}^{* (0)}_{j}    \mathcal{F}^{(0)}_k     ~~  . 
\end{align}
Lastly, $u^{(0)}_i$ is fully even under $x_i \rightarrow -x_i $, making the integrand of $I_3$ fully odd under $x_i \rightarrow -x_i $, and thus $I_3 =0$. Combining the results for $I_1$ and $I_2$, Eq.~\eqref{eq:reciprocalequations} narrows down to
\begin{align} \label{eq:reciprocalmaster}
           \mathcal{F}^{(1)}_j e^{*}_j   =  J^{(1)}_{jk}   
     \mathcal{F}^{* (0)}_{j}    \mathcal{F}^{(0)}_k   ~~ .  
\end{align}
We now consider two choices for $e^{*}_i$ which enable us to extract drag and lift force respectively. First, we take $e^{*}_i = e_i$. Using Eq.~\eqref{eq:simpleidentity}, Eq.~\eqref{eq:reciprocalmaster} then reduces to the formula for the convective correction to drag force \cite{Brenner_Cox_1963,masoud2019reciprocal,proudman1957expansions,vandyke1975perturbation}
 \begin{align}  \label{eq:reciprocalequations2}
    \begin{split} 
         &   \mathcal{F}^{(1)}_j   e_j   =    C_D^{(1)}     =  \frac{9 \pi }{4 }  + \mathcal{O}(\gamma_o^2 ) ~~   ,  \end{split}
\end{align}
Let us now consider the case $e^{*}_i = \varepsilon_{ij} e_j $. We find the convective correction to lift force to be given by
 \begin{align}  \label{eq:reciprocalequations12}
    \begin{split} 
         &  \varepsilon_{ij}       \mathcal{F}^{(1)}_i  e_j =    C_L^{(1)} = 
          \frac{9 \pi}{16 } \gamma_o   + \mathcal{O}(\gamma_o^2 )  ~~  .   \end{split}
\end{align}
Using Eq.~\eqref{eq:coefficients} and restoring the dimensionality, we finally obtain the total force
\begin{align}
\begin{split}
       F_i   =  & 6 \pi a  \bigg[ \eta_s   \left(1+ \frac{3}{8}  \text{Re} \right) \delta_{ij}   +  \frac{1}{2}  \eta_o  \left(1+ \frac{3}{16}  \text{Re} \right) \varepsilon_{ij}   \\  & 
 + \mathcal{O}(\text{Re}^2 ,\gamma_o^2 ) \bigg]       U^{\infty}_j  ~~ . 
 \end{split}
\end{align}

\section{Discussion}
In this work we studied the effect of convection for odd viscous flow past a sphere and computed low Reynolds number corrections to the corresponding forces and torques. Since a low Reynolds number expansion is a singular problem, we obtained a solution based on asymptotic matching of an inner and outer solution \cite{vandyke1975perturbation}. We considered two cases, namely the case where the vector $\ell_i$, which represents the intrinsic rotation of the odd viscous fluid is parallel and orthogonal to the free-stream velocity. For the longitudinal case, we could use axial symmetry to fully perform the asymptotic matching up to first order. Due to axial symmetry, lift force vanishes, however we do find that the interplay between convection and odd viscosity can turn on torque for a translating sphere. Torque does not arise for Stokes flow, because in this case a coupling between torque and translation is ruled out by the reflection symmetry of Stokes equation \cite{khain2023trading}. For the transverse case, the computation of convective effects on flow is more involved due to the absence of axial symmetry. A tool that was used to still be enabled to compute the first Reynolds number correction to Stokes lift is the odd generalization of the Lorentz reciprocal theorem \cite{hosaka2021nonreciprocal}. We only considered the case of lift force on a sphere, however it is known that the Lorentz reciprocal theorem allows for making general statements about forces on non-spherical obstacles \cite{Brenner_1961,Chester_1962,Brenner_Cox_1963,masoud2019reciprocal}. In particular, this was found to be true for the case where the obstacle is such that its shape alone cannot induce lift. For the case where the obstacle gives rise to lift for the case where there is only shear viscous flow, the term $I_3$ in Eq.~\eqref{eq:thirdterm} gives rise to a contribution to the lift force that requires full knowledge of the Stokes flow induced by the obstacle \cite{Brenner_Cox_1963}. For the same reason, it is not possible to find a shape-independent prediction for the first Reynolds number correction to odd viscosity induced lift force on an obstacle, even when the shape of this obstacle is such that it would not generate lift without odd viscosity. However, the Lorentz reciprocal theorem still allows efficiently computing this term for specific obstacle shapes for which $I_3$ vanishes by symmetry, as is the case for a spherical obstacle. We exclusively considered the first order odd viscous corrections to shear viscous flow, it would be interesting to see how our results generalize to the case of a non-perturbative odd viscosity, as has been done for the case of Stokes flow. \cite{PhysRevLett.132.218303}.

\section{Acknowledgements}
We thank Jeffrey Everts, Yuto Hosaka, Tali Khain, Pawe\l{} Matus, Colin Scheibner and Piotr Surówka for useful discussions.

\appendix

\onecolumngrid

\section{Computation of odd oseenlet}
\label{sec:oddoseen}
To derive the outer solution which obeys the inner boundary conditions Eq.~\eqref{eq:outerboundarycond}, we start from Eq.~\eqref{eq:fullequation1233214} and again add a point-force which approximates the sphere at distances far away from it. Introducing the point force yields the equations
\begin{subequations}  \label{eq:fullequat21213}
\begin{align} \label{eq:incopmressibility}
\tilde \nabla_i \tilde  u^{\prime (1)}_i   & =0 ~~ ,  \\   \label{eq:momentumbalace}
  (   \delta_{ij}   +  \gamma_{\text{o}} \varepsilon_{ij}  ) \tilde \Delta \tilde u^{\prime (1)}_j   -   \tilde \nabla_i \tilde p^{\prime (1)}_{\text{m}}  & =       e_j \tilde   \nabla_j \tilde  u^{ \prime (1)}_i  +    \mathcal{F}^{(0)}_i \tilde \delta     ~~  , 
\end{align}
\end{subequations}
where $\tilde \delta $ is the three-dimensional Dirac delta function located at the center of the sphere. Following Ref.~\cite{Pozrikidis1996-dq}, we take the divergence of Eq.~\eqref{eq:momentumbalace} to obtain
\begin{align} \label{eq:pressurepoisson}
  \tilde   \Delta \tilde p_{\text{m}}^{\prime (1)}    = -    \mathcal{F}^{(0)}_i 
 \nabla_i  \tilde \delta   +  \gamma_o \varepsilon_{ij} \tilde  \Delta  
\tilde  \nabla_i u^{\prime (1)}_j ~~  . 
\end{align}
Using the relation $\tilde \delta =   -  \tilde \Delta  \frac{1}{4 \pi \tilde r }  $, Eq.~\eqref{eq:pressurepoisson} can be solved as
\begin{align} \label{eq:pressuresolution}
     \tilde p^{\prime (1)}_{\text{m}}    =    \mathcal{F}^{(0)}_i  
\tilde  \nabla_i \frac{1}{4 \pi \tilde  r} +   \gamma_o \varepsilon_{ij}  \tilde  \nabla_i \tilde u^{\prime (1)}_j ~~ .  \end{align}
 Plugging this back into Eq.~\eqref{eq:momentumbalace}, we obtain
 \begin{align}  \label{eq:oseenequation21212}
  \begin{split}
    e_j \tilde \nabla_j  \tilde  u^{\prime (1)}_i     =    &  \left[  \delta_{ij}  \tilde  \Delta  + \gamma_{\text{o}}  ( \varepsilon_{ij} \tilde \Delta  -  \varepsilon_{k j } \tilde  \nabla_i \tilde  \nabla_k )  \right] \tilde   u^{\prime (1 )}_j    +     \frac{  \mathcal{F}^{(0)}_j}{4 \pi }  \left(  \delta_{ij} \tilde  \Delta   - \tilde  \nabla_i
\tilde  \nabla_j    \right) \frac{1}{\tilde r }    ~~ ,
   \end{split}  
\end{align}
At zeroth order in  $\gamma_{\text{o}} $, Eq.~\eqref{eq:oseenequation21212} can be solved with
\begin{align} \label{eq:oseenletansatz12}
    \tilde  u^{\prime (1,\text{s})}_i   = -    \mathcal{F}^{(0,\text{s})}_j   \left(  \delta_{ij}  \tilde  \Delta   -  \tilde  \nabla_i \tilde 
 \nabla_j    \right)  \mathbb{M}   ~~ . 
\end{align}
where $\mathbb{M}$ is given by
\begin{align}  \label{eq:oseenequation13}
 \left(  \tilde  \Delta   -       e_j \tilde  \nabla_j  
 \right) \mathbb{M}   =   \frac{1}{ 4 \pi \tilde r }       ~~ . 
\end{align}
Note that the ansatz of Eq.~\eqref{eq:oseenletansatz12} is such that Eq.~\eqref{eq:incopmressibility} is automatically satisfied. We can take the Laplacian of Eq.~\eqref{eq:oseenequation13} to find
\begin{align}  \label{eq:oseenequation15555}
 \left(  \tilde  \Delta   -       e_j \tilde  \nabla_j  
 \right) \Delta  \mathbb{M}   =   -\tilde \delta       ~~ . 
\end{align}
We define
\begin{align} \label{eq:definition}
   \tilde  \Delta \mathbb{M} = \Phi  \exp( \frac{\tilde x_i e_i }{2}  ) ~~ , 
\end{align}
so that Eq.~\eqref{eq:oseenequation15555} turns into the Helmholtz equation
\begin{align}
    \tilde \Delta \Phi  - \frac{1}{4} \Phi  = - \tilde \delta  ~~ , 
\end{align}
whose solution is given by \cite{Pozrikidis1996-dq}
\begin{align} \label{eq:phisolution}
    \Phi   =  \frac{1}{4 \pi \tilde r }  \exp(- \frac{\tilde  r}{2}  )  ~~ .
\end{align}
With Eq.~\eqref{eq:phisolution}, Eq.~\eqref{eq:definition} becomes
\begin{align} \label{eq:Mequation}
   \tilde    \Delta   \mathbb{M}   =   \frac{\exp(- \tilde\zeta   )}{ 4 \pi  \tilde r }  ~~ ,  
\end{align}
with $ \tilde \zeta =  
 \frac{1}{2} ( \tilde r  - \tilde  x_i e_i  ) $. Eq.~\eqref{eq:Mequation} can be solved to find \cite{Pozrikidis1996-dq}
\begin{align} \label{eq:Moutcome}
      \mathbb{M} = \frac{1 }{4  \pi   } \int_0^{\tilde \zeta }  \frac{1 - \exp(-  \xi)}{\xi} d \xi  ~~ . 
\end{align}
Having found $ \tilde u^{\prime (1,\text{s})}_i$, we move on to $\tilde u^{\prime (1,\text{o})}_i$. For this, we take the ansatz
\begin{align} \label{eq:oseenletansatz1}
  \tilde  u^{\prime (1,\text{o})}_i   =  -   \mathcal{F}^{(0,\text{s})}_j A_{i j k l } \tilde \nabla_k \tilde \nabla_l   \mathbb{N}   -    \mathcal{F}^{(0,\text{o})}_j   \left(  \delta_{ij}  \tilde  \Delta   -  \tilde  \nabla_i \tilde 
 \nabla_j    \right)  \mathbb{M}    ~~  , 
\end{align}
where $A_{i j k l }$ is some four-tensor. The second term on the right-hand side of Eq.~\eqref{eq:oseenletansatz1} cancels out the odd part of the point-force term in Eq.~\eqref{eq:oseenequation21212}. Plugging Eq.~\eqref{eq:oseenletansatz1} into Eq.~\eqref{eq:oseenequation21212} and considering the contributions $\mathcal{O} \left(\gamma_{\text{o}} \right)$ we find
\begin{align} \label{eq:expandedequation1}
\begin{split}
     &  \mathcal{F}^{(0,\text{s})}_j  A_{i j k l } \tilde \nabla_k \tilde \nabla_l 
 \left( \tilde \Delta \mathbb{N}  -   e_m \tilde \nabla_m   \mathbb{N}    \right)     =  - \mathcal{F}^{(0,\text{s})}_j   \left( \varepsilon_{il} \tilde \Delta  -  \varepsilon_{k l } \tilde  \nabla_i \tilde  \nabla_k \right)      \left(  \delta_{lj}  \tilde  \Delta   -  \tilde  \nabla_l \tilde 
 \nabla_j    \right) \mathbb{M}    ~~ . 
  \end{split}
 \end{align}
Simplifying leads to
\begin{align} \label{eq:expandedequation}
\begin{split}
     &  A_{i j k l } \tilde \nabla_k \tilde \nabla_l 
 \left(\tilde  \Delta \mathbb{N}  -   e_m \tilde \nabla_m   \mathbb{N}    \right)     =  -  \left( \varepsilon_{ij} \delta_{kl}
  + \varepsilon_{j k} \delta_{il}  - \varepsilon_{ i k } \delta_{l j}  \right) \tilde \nabla_k \tilde \nabla_l \Delta \mathbb{M}     ~~ . 
  \end{split}
 \end{align}
To solve Eq.~\eqref{eq:expandedequation}, we take
\begin{align}
  A_{ij k l }  =      \varepsilon_{ij} \delta_{kl}
  + \varepsilon_{j k} \delta_{il}  - \varepsilon_{ i k } \delta_{l j}  ~~ , 
 \end{align}
and also impose 
\begin{align} \label{eq:inbetweenequation}
    \tilde   \Delta \mathbb{N}  - e_i   \tilde \nabla_i   \mathbb{N}   = - \tilde  \Delta \mathbb{M} ~~  . 
 \end{align}
 Note that we have
 \begin{align}
 \begin{split}
        & \tilde \nabla_i \tilde u^{(1,\text{o})}_i = -  \tilde \nabla_i  \left[    \mathcal{F}^{(0)}_j  (  \varepsilon_{ij} \delta_{kl}
  + \varepsilon_{j k} \delta_{il}  - \varepsilon_{ i k } \delta_{l j} )  \tilde \nabla_k \tilde \nabla_l \mathbb{N} \right]  
    =    -  \frac{   \mathcal{F}^{(0)}_j }{\eta_s }  (  \varepsilon_{kj} \tilde \nabla_k
   + \varepsilon_{ j k } \tilde \nabla_k ) \tilde \Delta  \mathbb{N}  =0  ~~ , 
   \end{split}
 \end{align}
 and thus incompressibility is guaranteed for $u^{(1,\text{o})}_i $. Proceeding, we find that with the help of Eq.~\eqref{eq:Mequation}, Eq.~\eqref{eq:inbetweenequation} can be rewritten as
 \begin{align} \label{eq:newequation}
     \tilde  \Delta \mathbb{N}  -   e_i \tilde  \nabla_i   \mathbb{N}   =   -  \frac{\exp(- \tilde \zeta   )}{ 4 \pi \tilde r }  ~~ , 
 \end{align}
which is solved when one takes
\begin{align}
     \mathbb{N}  = \frac{\exp(- \tilde \zeta   )}{ 4  \pi    }~~ . 
 \end{align}
 The combination of Eqs.~\eqref{eq:pressuresolution},~\eqref{eq:oseenletansatz12} and~\eqref{eq:oseenletansatz1} yields Eq.~\eqref{eq:firstoseensolutionfull} in the main text.


\begin{thebibliography}{10}

\bibitem{Banerjee2017}
Debarghya Banerjee, Anton Souslov, Alexander~G. Abanov, and Vincenzo Vitelli.
\newblock Odd viscosity in chiral active fluids.
\newblock {\em Nature Communications}, 8, Nov 2017.

\bibitem{markovich2021odd}
Tomer Markovich and Tom~C. Lubensky.
\newblock Odd {{Viscosity}} in {{Active Matter}}: {{Microscopic Origin}} and {{3D Effects}}.
\newblock {\em Phys. Rev. Lett.}, 127(4):048001, 2021.

\bibitem{Fruchart_2023}
Michel Fruchart, Colin Scheibner, and Vincenzo Vitelli.
\newblock Odd viscosity and odd elasticity.
\newblock {\em Annual Review of Condensed Matter Physics}, 14(1):471–510, March 2023.

\bibitem{Khain_2022}
Tali Khain, Colin Scheibner, Michel Fruchart, and Vincenzo Vitelli.
\newblock Stokes flows in three-dimensional fluids with odd and parity-violating viscosities.
\newblock {\em Journal of Fluid Mechanics}, 934, January 2022.

\bibitem{PhysRevE.89.043019}
Matthew~F. Lapa and Taylor~L. Hughes.
\newblock Swimming at low reynolds number in fluids with odd, or hall, viscosity.
\newblock {\em Phys. Rev. E}, 89:043019, Apr 2014.

\bibitem{hosaka2023hydrodynamics}
Yuto Hosaka, Ramin Golestanian, and Abdallah Daddi-Moussa-Ider.
\newblock Hydrodynamics of an odd active surfer in a chiral fluid, 2023.

\bibitem{Kirkinis_Olvera}
E.~Kirkinis and M.~Olvera de~la Cruz.
\newblock Taylor columns and inertial-like waves in a three-dimensional odd viscous liquid.
\newblock {\em Journal of Fluid Mechanics}, 973:A30, 2023.

\bibitem{Lou2022}
Xin Lou, Qing Yang, Yu~Ding, Peng Liu, Ke~Chen, Xin Zhou, Fangfu Ye, Rudolf Podgornik, and Mingcheng Yang.
\newblock Odd viscosity-induced hall-like transport of an active chiral fluid.
\newblock {\em Proceedings of the National Academy of Sciences}, 119(42), October 2022.

\bibitem{PhysRevFluids.5.104802}
Alexander~G. Abanov, Tankut Can, Sriram Ganeshan, and Gustavo~M. Monteiro.
\newblock Hydrodynamics of two-dimensional compressible fluid with broken parity: Variational principle and free surface dynamics in the absence of dissipation.
\newblock {\em Phys. Rev. Fluids}, 5:104802, Oct 2020.

\bibitem{PhysRevE.90.063005}
Andrew Lucas and Piotr Sur\'owka.
\newblock Phenomenology of nonrelativistic parity-violating hydrodynamics in 2+1 dimensions.
\newblock {\em Phys. Rev. E}, 90:063005, Dec 2014.

\bibitem{berdyugin2019measuring}
A.~I. Berdyugin, S.~G. Xu, F.~M.~D. Pellegrino, R.~Krishna~Kumar, A.~Principi, I.~Torre, M.~Ben~Shalom, T.~Taniguchi, K.~Watanabe, I.~V. Grigorieva, M.~Polini, A.~K. Geim, and D.~A. Bandurin.
\newblock Measuring {{Hall}} viscosity of graphene's electron fluid.
\newblock {\em Science}, 364(6436):162--165, 2019.

\bibitem{Delacr_taz_2017}
Luca~V. Delacrétaz and Andrey Gromov.
\newblock Transport signatures of the hall viscosity.
\newblock {\em Physical Review Letters}, 119(22), November 2017.

\bibitem{KORVING19665}
J.~Korving, H.~Hulsman, H.F.P. Knaap, and J.J.M. Beenakker.
\newblock Transverse momentum transport in viscous flow of diatomic gases in a magnetic field.
\newblock {\em Physics Letters}, 21(1):5--7, 1966.

\bibitem{HULSMAN197053}
H.~Hulsman, E.J. {Van Waasdijk}, A.L.J. Burgmans, H.F.P. Knaap, and J.J.M. Beenakker.
\newblock Transverse momentum transport in polyatomic gases under the influence of a magnetic field.
\newblock {\em Physica}, 50(1):53--76, 1970.

\bibitem{tauber2019bulkinterface}
C.~Tauber, P.~Delplace, and A.~Venaille.
\newblock A bulk-interface correspondence for equatorial waves.
\newblock {\em J. Fluid Mech.}, 868:R2, 2019.

\bibitem{souslov2019topological}
Anton Souslov, Kinjal Dasbiswas, Michel Fruchart, Suriyanarayanan Vaikuntanathan, and Vincenzo Vitelli.
\newblock Topological {{Waves}} in {{Fluids}} with {{Odd Viscosity}}.
\newblock {\em Phys. Rev. Lett.}, 122(12):128001, 2019.

\bibitem{soni2019odd}
Vishal Soni, Ephraim~S. Bililign, Sofia Magkiriadou, Stefano Sacanna, Denis Bartolo, Michael~J. Shelley, and William T.~M. Irvine.
\newblock The odd free surface flows of a colloidal chiral fluid.
\newblock {\em Nat. Phys.}, 15:1188--1194, 2019.

\bibitem{han2021fluctuating}
Ming Han, Michel Fruchart, Colin Scheibner, Suriyanarayanan Vaikuntanathan, Juan~J. {de Pablo}, and Vincenzo Vitelli.
\newblock Fluctuating hydrodynamics of chiral active fluids.
\newblock {\em Nat. Phys.}, 17(11):1260, 2021.

\bibitem{hargus2020time}
Cory Hargus, Katherine Klymko, Jeffrey~M. Epstein, and Kranthi~K. Mandadapu.
\newblock Time reversal symmetry breaking and odd viscosity in active fluids: {{Green}}\textendash{{Kubo}} and {{NEMD}} results.
\newblock {\em J. Chem. Phys.}, 152(20):201102, 2020.

\bibitem{PhysRevLett.130.158201}
Anthony~R. Poggioli and David~T. Limmer.
\newblock Odd mobility of a passive tracer in a chiral active fluid.
\newblock {\em Phys. Rev. Lett.}, 130:158201, Apr 2023.

\bibitem{PhysRevLett.127.178001}
Cory Hargus, Jeffrey~M. Epstein, and Kranthi~K. Mandadapu.
\newblock Odd diffusivity of chiral random motion.
\newblock {\em Phys. Rev. Lett.}, 127:178001, Oct 2021.

\bibitem{matus2024molecular}
Paweł Matus, Ruben Lier, and Piotr Surówka.
\newblock Molecular modelling of odd viscoelastic fluids, 2024.

\bibitem{Tan2022}
Tzer~Han Tan, Alexander Mietke, Junang Li, Yuchao Chen, Hugh Higinbotham, Peter~J. Foster, Shreyas Gokhale, J{\"o}rn Dunkel, and Nikta Fakhri.
\newblock Odd dynamics of living chiral crystals.
\newblock {\em Nature}, 607(7918):287--293, Jul 2022.

\bibitem{avron1995viscosity}
J.~E. Avron, R.~Seiler, and P.~G. Zograf.
\newblock Viscosity of {{Quantum Hall Fluids}}.
\newblock {\em Phys. Rev. Lett.}, 75(4):697--700, 1995.

\bibitem{avron1998odd}
J.~E. Avron.
\newblock Odd {{Viscosity}}.
\newblock {\em J. Stat. Phys.}, 92(3/4):543--557, 1998.

\bibitem{levay1995berry}
P{\'e}ter L{\'e}vay.
\newblock Berry phases for {{Landau Hamiltonians}} on deformed tori.
\newblock {\em J. Math. Phys.}, 36(6):2792--2802, 1995.

\bibitem{hoyosreview}
Carlos Hoyos.
\newblock Hall viscosity, topological states and effective theories.
\newblock {\em International Journal of Modern Physics B}, 28(15):1430007, 2014.

\bibitem{PhysRevLett.108.066805}
Carlos Hoyos and Dam~Thanh Son.
\newblock Hall viscosity and electromagnetic response.
\newblock {\em Phys. Rev. Lett.}, 108:066805, Feb 2012.

\bibitem{olveira}
Hang Yuan and Monica Olvera de~la Cruz.
\newblock Stokesian dynamics with odd viscosity.
\newblock {\em Phys. Rev. Fluids}, 8:054101, May 2023.

\bibitem{hosaka2023lorentz}
Yuto Hosaka, Ramin Golestanian, and Andrej Vilfan.
\newblock Lorentz reciprocal theorem in fluids with odd viscosity.
\newblock {\em Phys. Rev. Lett.}, 131:178303, Oct 2023.

\bibitem{lier2022lift}
Ruben Lier, Charlie Duclut, Stefano Bo, Jay Armas, Frank J{\"u}licher, and Piotr Sur{\'o}wka.
\newblock Lift force in odd compressible fluids.
\newblock {\em arXiv:2205.12704}, 2022.

\bibitem{dewit2023pattern}
Xander~M. de~Wit, Michel Fruchart, Tali Khain, Federico Toschi, and Vincenzo Vitelli.
\newblock Pattern formation by non-dissipative arrest of turbulent cascades, 2023.

\bibitem{lorentzoriginal}
Lorentz~H. A.
\newblock Eene algemeene stelling omtrent de be- weging eener vloeistof met wrijving en eenige daaruit afgeleide gevolgen.
\newblock {\em Versl. Kon. Acad. Wet.}, 5:168--175, 1896.

\bibitem{lier2024slipinduced}
Ruben Lier.
\newblock Slip-induced odd viscous flow past a cylinder, 2024.

\bibitem{hosaka2021nonreciprocal}
Yuto Hosaka, Shigeyuki Komura, and David Andelman.
\newblock Nonreciprocal response of a two-dimensional fluid with odd viscosity.
\newblock {\em Phys. Rev. E}, 103(4):042610, 2021.

\bibitem{PhysRevE.109.044126}
Charlie Duclut, Stefano Bo, Ruben Lier, Jay Armas, Piotr Sur\'owka, and Frank J\"ulicher.
\newblock Probe particles in odd active viscoelastic fluids: How activity and dissipation determine linear stability.
\newblock {\em Phys. Rev. E}, 109:044126, Apr 2024.

\bibitem{PhysRevE.104.064613}
Yuto Hosaka, Shigeyuki Komura, and David Andelman.
\newblock Hydrodynamic lift of a two-dimensional liquid domain with odd viscosity.
\newblock {\em Phys. Rev. E}, 104:064613, Dec 2021.

\bibitem{PhysRevLett.132.218303}
Jeffrey~C. Everts and Bogdan Cichocki.
\newblock Dissipative effects in odd viscous stokes flow around a single sphere.
\newblock {\em Phys. Rev. Lett.}, 132:218303, May 2024.

\bibitem{proudman1957expansions}
Ian Proudman and J.~R.~A. Pearson.
\newblock Expansions at small {{Reynolds}} numbers for the flow past a sphere and a circular cylinder.
\newblock {\em J. Fluid Mech.}, 2(3):237--262, 1957.

\bibitem{vandyke1975perturbation}
Milton Van~Dyke.
\newblock {\em Perturbation Methods in Fluid Mechanics}.
\newblock {Parabolic Press}, {Stanford, California}, 1975.

\bibitem{veysey2007simple}
John Veysey and Nigel Goldenfeld.
\newblock Simple viscous flows: {{From}} boundary layers to the renormalization group.
\newblock {\em Rev. Mod. Phys.}, 79(3):883--927, 2007.

\bibitem{lamb1932hydrodynamics}
Horace Lamb.
\newblock {\em Hydrodynamics}.
\newblock {Cambridge University Press}, {New York, U.S.A.}, 1932.

\bibitem{an1889second}
Whitehead AN.
\newblock Second approximation to viscous fluid motion: a sphere moving steadily in a straight line.
\newblock {\em QJ Pure Appl Math}, 23:143--152, 1889.

\bibitem{oseen1910uber}
Carl~Wilhelm Oseen.
\newblock \"uber die {{Stokes}}'sche {{Formel}}, und \"uber eine verwandte {{Aufgabe}} in der {{Hydrodynamik}}.
\newblock {\em Ark. Mat., Astron. Fys.}, 6:1, 1910.

\bibitem{khain2023trading}
Tali Khain, Michel Fruchart, Colin Scheibner, Thomas~A. Witten, and Vincenzo Vitelli.
\newblock Trading particle shape with fluid symmetry: on the mobility matrix in 3d chiral fluids, 2023.

\bibitem{markovich2022non}
Tomer Markovich and Tom~C. Lubensky.
\newblock Non reciprocal odd viscosity: Coarse graining the kinetic energy and exceptional instability, 2022.

\bibitem{ganeshan2017odd}
Sriram Ganeshan and Alexander~G. Abanov.
\newblock Odd viscosity in two-dimensional incompressible fluids.
\newblock {\em Phys. Rev. Fluids}, 2(9):094101, 2017.

\bibitem{chester_breach_proudman_1969}
W.~Chester, D.~R. Breach, and Ian Proudman.
\newblock On the flow past a sphere at low reynolds number.
\newblock {\em Journal of Fluid Mechanics}, 37(4):751–760, 1969.

\bibitem{kaplun1957low}
Saul Kaplun.
\newblock Low {{Reynolds Number Flow Past}} a {{Circular Cylinder}}.
\newblock {\em Indiana Univ. Math. J.}, 6(4):595--603, 1957.

\bibitem{Kimkarrila2013-op}
Sangtae Kim and Seppo~J Karrila.
\newblock {\em Microhydrodynamics: Principles and Selected Applications}.
\newblock Butterworth-Heinemann, Oxford, England, September 2013.

\bibitem{Pozrikidis1996-dq}
C~Pozrikidis.
\newblock {\em Introduction to theoretical and computational fluid dynamics}.
\newblock Oxford University Press, New York, NY, September 1996.

\bibitem{happel1983low}
John Happel and Howard Brenner.
\newblock {\em Low {{Reynolds}} Number Hydrodynamics}.
\newblock {Springer}, {The Hague, The Netherlands}, 1983.

\bibitem{masoud2019reciprocal}
Hassan Masoud and Howard~A. Stone.
\newblock The reciprocal theorem in fluid dynamics and transport phenomena.
\newblock {\em J. Fluid Mech.}, 879:P1, 2019.

\bibitem{Brenner_Cox_1963}
H.~Brenner and R.~G. Cox.
\newblock The resistance to a particle of arbitrary shape in translational motion at small reynolds numbers.
\newblock {\em Journal of Fluid Mechanics}, 17(4):561–595, 1963.

\bibitem{Brenner_1961}
Howard Brenner.
\newblock The oseen resistance of a particle of arbitrary shape.
\newblock {\em Journal of Fluid Mechanics}, 11(4):604–610, 1961.

\bibitem{Chester_1962}
W.~Chester.
\newblock On oseen’s approximation.
\newblock {\em Journal of Fluid Mechanics}, 13(4):557–569, 1962.

\end{thebibliography}
\end{document}